\documentstyle[epsfig]{mn}

\title[Inferring dark halo structure]{Inferring dark halo structure from observed scaling laws of 
late type galaxies and LSB's}
\author[X. Hernandez and G. Gilmore]{X. Hernandez$^{1,2}$ and Gerard. Gilmore$^1$ \\
$^1$ Institute of Astronomy, Cambridge University, Madingley Road, Cambridge CB3 0HA \\
$^2$ Instituto de Astronom\'\i a, Universidad Nacional Aut\'onoma de M\'exico, A.P. 70-264, 04510 M\'exico, D.F.}

\date{\today}

\begin{document}
\maketitle

\begin{abstract}
We re-examine the Fall \& Efstathiou (1980) scenario for galaxy formation,
including the dark halo gravitational reaction to the formation of the baryon
disk, as well as continuous variations in the intrinsic halo density profile.
The recently published rotation curves of LSB and dwarf galaxies together with 
previously known scaling relations provide sufficient information on the 
present day structure of late type disk galaxies to invert the problem. By requiring
that the models reproduce all the observational restrictions we can fully constrain
the initial conditions of galaxy formation, with a minimum of assumptions, in particular
without the need to specify a CDM halo profile. This allows one
to solve for all the initial conditions, in terms of the halo density profile, 
the baryon fraction and the total angular momentum. We find that a unique initial halo shape is sufficient
to accurately reproduce the rotation curves of both LSB and normal late type
spiral galaxies. This unique halo profile differs substantially from that found in standard CDM
models (Navarro, Frenk \& White 1996). A galactic baryon fraction of 0.065 is found. The initial value
of the dimensionless angular momentum is seen to be the principal discriminator between
the galaxy classes we examine. The present day scalings between structural parameters
are seen to originate in the initial conditions.  
\end{abstract}

\begin{keywords} 
galaxies: formation --  galaxies: structure -- dark matter
\end{keywords}

\section{Introduction}

A common approach to the study of galaxy formation is to start from
a set of initial conditions, defined as a self consistent cosmological
model including the phase distribution and amplitude of an initial
density fluctuation spectrum, and from this to construct a full
realization of the growth of structure.
The gravitational evolution of the early fluctuations are typically followed
numerically, so that gravitationally bound structures can be identified,
and depending on the spatial resolution of the numerical code, traced over different
scales. The success of this method in reproducing the large scale
structure of the universe is well known (e.g. Davis et al. 1985, White et al. 1987). The
problem of going to galactic scales is complicated by the need to 
include gas dynamics, star formation rates, energy feedback between forming
and exploding stars and the other baryonic components, stellar initial mass functions,
and other physical processes pertaining to the baryonic component,
which are only marginally understood. This can be seen most clearly in
the inability of this approach to reproduce the constant density core
structure seen in the dark halos of both dwarf and LSB galaxies (see
for example the rotation curves of de Blok et al. (1996) for LSB's,  
Burkert (1995) for dwarf galaxies, or Carignan \& Beaulieu (1989) for 
DD0 154, Pryor \& Kormendy (1990) for dSph density profiles, and Ibata et al. (1997) 
for dynamical evidence for core structure in
dSph galaxies, and Burkert \& Silk (1997) for a discussion of this point). 
  
The spirit of this work is to derive as much information on the
initial conditions of galactic formation as can be obtained from the
available observations of the galaxies themselves, with the smallest number
of assumptions. Uncertainties in the
interplay between star formation, gas dynamics and other non linear
processes of galactic formation
will necessarily limit any simple analysis, such as the one made here.
There is the advantage of being able to derive information
about the present state of galaxies and the
scaling laws they follow with minimal dependence on an adopted primordial structure of the universe.
Taking an approach that starts from the present day structure of galaxies rather 
than from the initial conditions will severely limit the scope of the predictions
of our work. It is only through restricting the free parameters to the minimum number (one)
and understanding directly how uncertainties propagate through the model, that we retain
predictive power, rather than simply reproduce the observations which we introduce as
constraints.

Present day late type galaxies come in a variety of sizes and shapes,
but appear to obey well established scaling laws. To a first
approximation, disk galaxies can be described by  four observational
parameters: the total luminosity of the exponential light surface
density profile and the radial scale length of this profile, the
central density and the asymptotic rotation velocity of the dark matter
halo. The first two of these parameters relate directly to the baryonic
component, and the last two refer mostly to the dark matter. The total luminosity
(mass?) is tightly correlated with the asymptotic rotation velocity through the
Tully-Fisher relation (Tully \& Fisher 1977), the total luminosity also scales with the
exponential disk scale length (Flores et al. 1993). The question addressed in this work is
whether the observationally available information on present day
galactic structural parameters and their scaling laws is sufficient to
introduce some constraints on the initial conditions of these galaxies,
without having to impose some specific dark matter density profile.
To this end we apply a simple method of mapping a set of initial conditions in
terms of virialized halo structure, baryon fraction and angular
momentum into a set of present day galactic structural
parameters. This method is essentially the one introduced by Blumenthal et al. (1986)
and Flores et al. (1993), which we extend here to the case of a halo
density profile which is treated as a model parameter. 
We restrict analysis to late-type and LSB galaxies, as only in these cases is the
influence of dark matter substantial at all radii. Analysis of earlier type
galaxies is necessarily more sensitive to the evolution of the baryonic components.
We constrain
all the initial conditions through the condition that the resulting
galactic structure follows the observational constraints, taking
special note of the central density of the dark halos, which can be
derived from the recently published rotation curves of LSB galaxies of
de Blok et al. (1996). Once this is done, we wish to investigate what
relations between the initial conditions result from having imposed
certain observational restrictions. It is possible that when seen in
terms of initial conditions, observational laws such as the
Tully-Fisher relation or the approximate scaling of the disk mass with
the square of the scale radius, appear natural. Additionally, one
is interested to see to what extent different galaxy classes populate
different regions in the initial condition space. 
As no specific dark halo shape is imposed, we use present day galactic structure to
solve for the halo profile, which we find to differ from that which results from
the fully self-consistent structure formation CDM simulations of 
Navarro, Frenk \& White (1996), henceforth NFW. A comparison of this
two dark halo profiles will be made, in connection to the well studied dark halo
of a particular gas rich dwarf galaxy, DDO 154. 

Section (2) presents a dimensional analysis to the problem, which is
expanded in section(3) to a more complete galactic formation
scenario, calibrated using the data on LSB galaxies. In
Section (4) we apply the previous framework to normal late type
galaxies, and in section (5) we present the conclusions of this work.

\section{Analytical approximation}

Although most of the material in this section can be found elsewhere, it is included here to give the 
reader an intuitive framework for the problem, and to present a zero
order solution to it. The three galactic structural parameters which
are most readily accessible to observations in late type galaxies are $M_d$,
the total disk mass, $R_d$, the disk exponential scale radius and
$V_M$, the maximum rotation velocity. $M_d$ is derivable from total luminosities once a $M/L$
ratio has been assumed. Uncertainties in the value of this ratio are
generally of less than a factor of 2 for a given system. These three
quantities are seen to be correlated through the Tully-Fisher (T-F) law $M_d \propto V_M ^\alpha$, and
an approximate scaling of $M_d \propto R_d ^\beta$, with $\alpha$ close to 4 and 
$\beta$ close to 2. In the following section we present data from the literature
yielding $\alpha =3.5, \beta = 2.33$. It is perhaps easier to
understand these correlations if one translates the observable
quantities into initial conditions for the galaxy, for example, a
total mass $M$, total potential energy $W$, baryon fraction $F$,
and a total angular momentum, $L$. Once $M$ and $F$ are
fixed, the total baryon content is given by $FM$. It is customary to
refer not to $L$, but to the dimensionless quantity 
$$
 \lambda = {L|{W \over 2}|^{1/2} \over G M^{5/2}}, 
$$
where $G$ is Newton's constant, so it shall be treated thus in this
work. On purely dimensional grounds, we can establish the following relations
between the initial conditions and the observed parameters:

\begin{equation}
M_d =F M; V_M  \propto \left( W \over M \right)^{1/2} ; R_d \propto {\lambda M^2
\over W}. 
\end{equation}

at fixed $F$ and $\lambda$ the T-F law translates into
a relation between $M$ and $W$. Substituting $W$ for $R_d$ from the above
relations one obtains:

\begin{equation}
M \propto M_d \propto R_d ^{\alpha \over \alpha -2}  
\end{equation}

i.e. for a galaxy population having constant $F$ and $\lambda$, which
could represent a fixed galaxy-type sample, one expects a connection
between the T-F index, and the exponent in the $M_d vs. R_d$
relation. For $\alpha =4$ equation (2) gives $\beta =2$, quite in
agreement with observations. 

Using the above relations between the
initial conditions and the final result to translate the observational
relations one obtains:

\begin{equation}
{M^{2\beta -1} \over F} \propto \left( W \over \lambda \right)^ \beta;
FM^{{\alpha +2} \over 2} \propto  W^{\alpha /2}  
\end{equation}

Eliminating $M$ and $W$ from the above two relations for $\beta=\alpha /(\alpha -2) $, we get:

\begin{equation}
F \propto  \lambda 
\end{equation}

Equation (4) implies that for a population of galaxies following the
two observational relations $\alpha$ and $\beta$ constant, the baryon
fraction must scale with the dimensionless spin parameter. This result
is compatible with what was found by Flores et al. (1993),
and essentially states the ``conspiracy'' between angular momentum and
matter content, responsible for the observed flat rotation
curves. Additionally, one expects $\lambda \approx 0.05$ and $F \approx 0.05$.
It is interesting that the condition $\lambda \approx F$
actually holds for any system where a mass $FM$
orbits in centrifugal equilibrium in the potential generated by $M$,
which might be a clue as to the origin of this relationship.

Taking the first of relations (3) at constant $F$ and $\lambda$, we obtain

\begin{equation}
W \propto M^{{2\beta -1}\over \beta},
\end{equation}

which is equivalent to the initial conditions translation of the T-F,
given equation (2). If we now translate $W$ into a typical formation
redshift (collapse z) of the system through

\begin{equation}
(1+z) \propto {W \over M^{5/3}} =C \left( {W \over M^{5/3}} \right),  
\end{equation}

valid for a spherical ``top hat'' fluctuation with no initial internal kinetic energy
prior to virialization, in an $\Omega =1, \Lambda
=0$ universe, equation(5) gives:

\begin{equation}
(1+z)^{3\beta \over {\beta-3}} \propto M. 
\end{equation}

For $\beta=2$ equation(7) gives $(1+z)^{-6} \propto M$, which is what results
from a simple Press-Shechter analysis for a spectral index in the
galactic region of $n=-2$, from:

\begin{equation}
(1+z)^{-6 \over n+3} \propto M.  
\end{equation}

Even though this previous analysis has been simplified to the extreme,
it suffices to show that the observational relations $\alpha$ and
$\beta$ constant are not independent, and are the result of the initial conditions of galactic
formation, related to the structure formation scenario. Much
of what has been presented here can be found in several sources,
amongst them Peebles (1993), White (1997), Padmanabhan (1993). Equation (8) was taken from Padmanabhan (1993).
The following section introduces a more detailed way of relating the
observed structural parameters to the initial conditions of galaxy formation.

\section{Theoretical approach}

In this section we describe the procedure which we use to associate a set of initial formation conditions
to a present day galaxy. We are assuming that the galaxies we are modeling passed through a phase in which
they were characterized by an initial halo configuration, in which the baryon fraction was mixed 
homogeneously with the dark matter in a virialized configuration, having a certain total mass, total
potential energy, and total angular momentum. At this stage baryonic cooling sets in and leads to the 
formation of the galactic disk. The initial halo is taken to have spherical symmetry, and not 
to suffer any further accretion or merging processes. The formation of this initial halo is not treated here.
Whether it formed through gradual radial accretion, through a complicated merging history or was characterized
by a monolithic structure throughout its history is not relevant to our study. Any one of the above 
scenarios is compatible with our assumptions. The real requirement is that the dark halo was fully formed at the time
the bulk of the baryons begin to cool and fall into the center of the dark halo, which
seems a reasonable assumption if we
restrict our attention to gas-rich disk galaxies. The dynamical frailty of the disks and the non dissipative
nature of stars suggest that no major mergers have taken place since the disks formed, and that they were 
assembled from gas.   

To fully characterize this initial configuration, for every galaxy one requires five
pieces of information, {\it viz.} the choice of a halo density profile, a
total mass, $M$, a total potential energy, $|W|$, a baryonic mass
fraction, $F$, and a value of $\lambda$.
Whereas there is mounting evidence, both theoretical and observational which
suggests the existence of a universal shape for the density profile of
dark halos, it is not clear what this shape should be. This curious
situation arises from the fact that the theoretically predicted shapes
for dark halos do not correspond to the observed ones 
(Burkert 1995, Burkert \& Silk 1997). It is not clear
at this point if the original shapes of the dark halos were the ones
predicted by CDM theory, which were later modified by secular processes (Navarro et al.  1996),
or whether they always had the observed shapes, there being some extra
physics yet to be considered by CDM theory. Not wanting to impose upon the
model a specific halo shape {\it a priori}, we assume only that the
initial dark halos can be characterized by a King profile. King
profiles form an infinity of families of density profiles, determined
by a shape parameter, $P_0= \Phi(0) / \sigma^2 $, the value of the
gravitational potential at the center of the distribution, in units of
the velocity dispersion. For every value of $P_0$ there exists a
family of King density profiles, having two independent parameters, a
central density and a core radius, or alternatively, a total mass and
a total potential energy. Varying these three parameters, one can
obtain virtually any density distribution with a radially increasing
negative logarithmic slope, and a central behavior shallower than
$r^{-2}$. Thus, King distributions have a finite total mass and can
naturally accommodate a constant density inner region, such as
characterizes observed dark matter halos. We show below they describe galaxies
at least as well as does the limiting case King profile, a (standard) lowered 
isothermal profile. Even though King
distributions also include a self-consistent distribution function,
and have been found to represent well the end product of some N-body
relaxation processes, we are not requiring here that the initial dark halos
were strictly King profiles, only that their density profiles could be
approximated reasonably well by King profiles. (Recall that the
mathematical process of obtaining a distribution function from a given
density profile does not yield a unique answer in general.) Requiring
that our initial halos were King profiles is hence a much weaker
imposition than choosing some two-parameter halo model such as a
Hernquist, lowered isothermal, Hubble, or other, as is often done in
similar work (Dalcanton et al. 1996, Flores et al. 1993). We keep the assumption of
self similarity for dark matter halos, but treat the actual profile
shape, $P_0$ as a parameter to be fixed by the observational data, in
keeping with the aims of this work.

Once a set of $M, W, P_0, F$ and $\lambda$ are chosen to represent
the initial conditions, we turn a fraction $F$ of the
original halo into a disk at the center of the halo, having a radial surface density 
profile given by

\begin{equation}
\Sigma(R)=\Sigma_0 e^{-R/R_d}  
\end{equation}

The two parameters in equation (9) are fixed by the two conditions of
conservation of mass and angular momentum. The baryonic
disk was assumed to be in centrifugal equilibrium, and no angular
momentum transfer between the baryons and the dark halo was
considered. The dark halo now has mass $M(1-F)$, and a structure
determined by the condition 
$$
R_{i}M_{i}(R_i)=R(M_{H}(R)+M_{d}(R)),
$$
where $R_{i}$ is the orbital radius of a dark matter particle orbiting
within the original virialized distribution $M_{i}(R_i)$, $R$ the
final radius of this particle after the disk has formed, $M_{d}(R)$
the final disk mass distribution, and $M_{H}(R)$ the final halo
structure. In this way, assuming the disk formation process is not
violent, we use an adiabatic invariant for the halo particles to
calculate the final halo structure 
(see Flores et al. 1993, Firmani et al. 1996, Dalcanton et al. 1996).

It has been known since the first time baryonic infall models were calculated
(Barnes \& Efstathiou 1987) that the adiabatic invariant approximation breaks down in the
innermost regions. In that work it is shown that this approximation is not valid in the
inner 1-2 disk scale lengths where more complicated dynamics take over, 
for the normal late type galaxies they treated. However,
at this point we are modeling LSB galaxies, which clearly have undergone much milder dissipative
processes, the baryons have contracted much less, and the halo reaction has been more limited.
This improves the validity of the adiabatic invariant approach into the central regions 
significantly, as can be seen from Figure (1), even the central regions of LSB
galaxies have suffered little dissipation. When we come to normal late type spirals, we are 
aware that the adiabatic approximation is not valid in the central regions, where baryon dissipation
has been sufficiently strong to actually make that component dynamically dominant in the inner regions.
This last point also introduces other uncertainties, as the high density of baryons
implies that gas hydrodynamics and the dynamical coupling between the stellar and gaseous
components through star formation, stellar winds and supernovae, might prove the dominant
processes. In this way, the inner 1-2 disk scale lengths of the rotation curves we calculate
for normal late type galaxies are probably not relevant. In spite of this, out conclusions
are not affected, as it is only in the LSB cases that we use the inner regions of the 
rotation curves to compare with observation.

At this point, we have the complete baryon disk and dark matter halo
matter distributions, and therefore the complete rotation curve as
well. For simplicity only the monopole term for the disk potential was
calculated, such as is done in other similar work (Dalcanton et al. 1996). See
Figure (2-17) in Binney \& Tremaine (1987) where this approximation is compared to
the full rotation curve of a totally flat exponential disk. The
agreement would improve for an exponential disk with some vertical
structure, such as one expects in reality. 

In this work we are only using the fact that the final baryonic
matter disk has an exponential profile, as deduced from observations of
late type disks. The physics which leads to this profile is of no consequence to us
here so long as the adiabatic invariant assumption applies. 
More detailed galactic disk evolutionary studies have addressed
this point: e.g. Struck-Marcell (1991) finds that the
exponential profile is a hydrostatic equilibrium configuration in an
isothermal halo; Sayo \& Yoshi (1990) obtain this profile as a consequence
of the viscous dissipation mass and angular momentum transfer processes under
the assumption of a viscous timescale similar to the star formation
timescale; or the more complete Firmani et al. (1996), who consider a
self consistent galactic disk evolutionary scenario including
independent estimates of star formation rates and viscous timescales,
and find the exponential profile as a natural result of the problem,
independent of the initial disk profile. Whatever the details, the slower the
disk forms, the better suited the adiabatic invariant treatment for
the halo becomes. The exponential disk profile and accompanying halo
profile are intended to represent the present situation to be compared
with observations, and not the initial disk structure.

With the method outlined above, we translate a set of initial
conditions $M, W, P_0, F$ and $\lambda$ into a set of present day
observable conditions, $V_M, \rho_0, R_M, M_d$ and $R_d$, the maximum
rotation velocity, the final total central matter density, a measure
of the total extent of the dark halo, the total disk
mass, and the disk scale length, respectively. $R_M$ was defined as the radius
at which the rotation velocity has decreased to 85\% of its maximum value, 
$0.85 V_M$, a value yet to be observed in most cases (notice however the case
of DDO 154, which will be discussed more fully later). The total disk
mass was compared to the observed total luminosities adopting $M/L$
ratio of 2 in the B band for gas rich LSB
systems, and disk galaxies with types $>4$. 
All observations were scaled to $H_0 =75 kms^{-1}Mpc^{-1}$
where necessary. This value of $H_0$ is used throughout, which makes the value
of the constant in eq.(6) $C=7.536\times 10^{36}$.
Units throughout are $M_{\odot}$ and kpc, $ G=4.45\times10^{-39} kpc^3 M_{\odot}^{-1} s^{-1}$, 
rotation velocities are given in
$km s^{-1}$. 

\begin{figure}
\epsfig{file=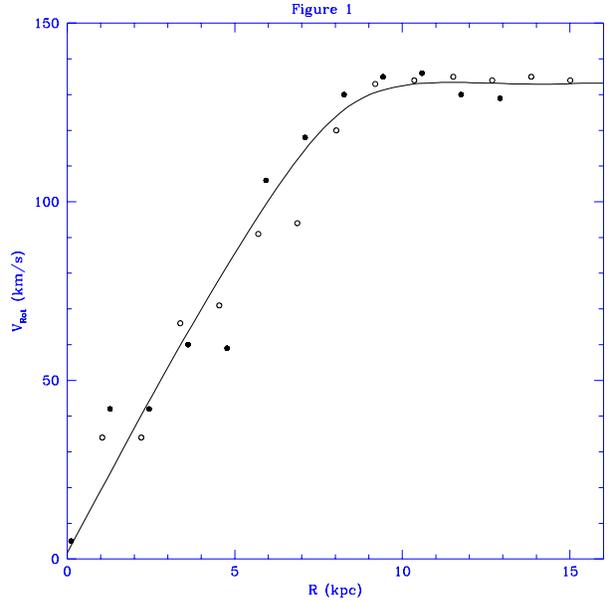,angle=0,width=8.5cm}
\caption{Rotation curve of one of the LSB galaxies in the de Blok et al. (1996)
sample (F571-8), from their data. Notice the clearly defined central core region, 
with a linear relation between rotation velocity and radius, which has been used to
obtain $\rho_0$ for this systems.}
\label{Figure 1.}
\end{figure}

Having the initial conditions specified as a 5 dimensional vector, we
require 5 restrictions on any galaxy we want to model to determine
what its initial conditions were. The first case we model is that of
the newly observed LSB galaxies, which are the systems for which the
most information can be obtained. In most disk galaxies the baryonic
component is gravitationally dominant in the central regions, which
makes estimating the total central matter density a difficult problem,
necessarily sensitive to the assumed $M/L$ ratios and orbital anisotropy.
In the case of the LSB galaxies, the
large disk scale radius yields a very extended baryonic component
(presumably arising from a high initial angular momentum) such that
these systems are dark matter dominated into their innermost
regions. Given this, from measurements of the rotation curves a good
estimate of the total central density is available, see Figure
(1). Despite this situation, the only available sample of LSB rotation
curves (de Blok et al. 1996) contains only a small sample of 
galaxies (16), such that it is hard to distinguish any trend 
in the LSB galaxy-type (after all, this
is a highly empirical definition, resting merely on the observational
detectability limits, which do not necessarily correlate with galactic
types. It has been pointed out that these are very new measurements, and
that many more are expected as the observational situation improves).  
We further emphasize that the central rotation curves remain not well sampled,
and may be further degraded by beam smearing and other finite resolution effects.
Nonetheless, fitting straight lines to the inner regions of the rotation curves
published by de Blok et al. (1996), we can estimate the total central matter
densities of these systems, a plot of which is presented in Figure
(2). As considerable spread is evident (at least some of it observational), and only galaxies within a
narrow range of disk masses are present, rather than attempting to
extract a trend from these galaxies, we only took the median of the
distribution to characterize a ``typical'' LSB, in terms of a total baryon
mass and total central matter density of $log(M_d)=9.5$, $log(\rho_{0}/M_{\odot} pc^{-3})=-1.85$.
These will be our first two
observational restrictions, used to construct a model of a typical LSB
galaxy.

\begin{figure}
\epsfig{file=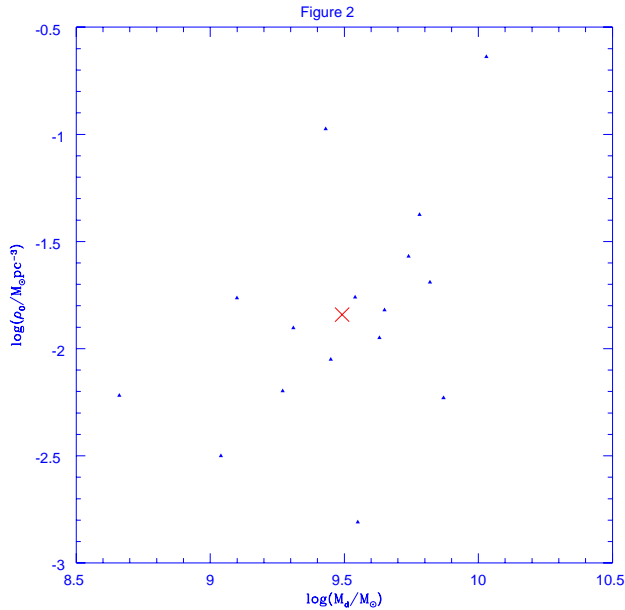,angle=0,width=8.5cm}
\caption{Plot of $\rho_0 vs. M_d$ for the galaxies in the de Blok
sample. Given the large dispersion, only the mean values of $\rho_0$ and $M_d$ for a typical
LSB galaxy were used, shown by the cross.}
\label{Figure 2.}
\end{figure}

From observations of total luminosity and rotation velocity
amplitude it appears that LSB galaxies
follow the same Tully-Fisher relation as normal late type spirals
e.g. Zwaan et al. (1995). We
take the recent Tully-Fisher law determination of Rhee \& Van Albada (1995),
where a population corrected total disk mass is calculated from colour
information to derive a disk mass vs. maximum rotation velocity
Tully-Fisher relation. That study shows that by taking into account
the different $M/L$ ratios associated with different galaxies, the
dispersion in the Tully-Fisher relation can be reduced to a
minimum. The resulting slope in $M_d vs. V_M$ is quite similar to that of the I band 
Tully-Fisher relation, but the dispersion goes down by a factor of 2.
Given the greater ease with which this can be compared to our
study, we use this Mass Tully-Fisher throughout this work, valid for all galaxies, 

\begin{equation} 
log(M_d)=2.8+3.5 log(V_M). 
\end{equation}

This T-F relation provides
a third constraint for our typical LSB.

\begin{figure}
\epsfig{file=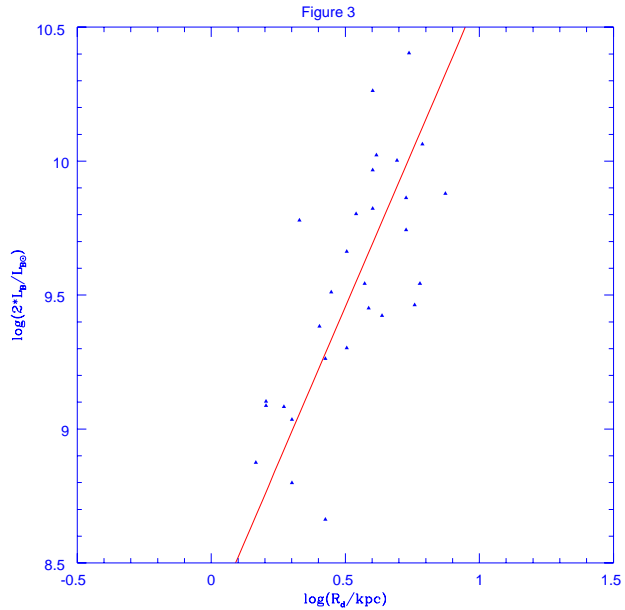,angle=0,width=8.5cm}
\caption{Log ($M_d$) vs. log($R_d$) for the LSB galaxies from the samples of 
McGaugh \& Bothun (1994) and Sprayberry \& Impey (1995), assuming a $M_{\odot}/L_{B\odot}$ ratio of 2.
 Also shown is a linear fit to the data.
The slope of this line is what equation(2) predicts, for the T-F law of Rhee \& van Albada (1995).}
\label{Figure 3.}
\end{figure}

Combining the samples of McGaugh \& Bothun (1994) and Sprayberry \& Impey (1995), who have studied
the light distribution in LSB galaxies, we compiled a list of LSB
galaxies for which a measured total luminosity and luminosity exponential
scale radius exist. These data are displayed in Figure (3). A very
clear trend is apparent with a linear regression fit giving 

\begin{equation}
log(M_d)=8.31+2.27 log(R_d). 
\end{equation}

The variance in the slope is of 0.17, and of 0.05 in the intercept.
This relation is assumed to hold for all
the LSB galaxies and provides the fourth observational
constraint on the typical LSB we want to model. Notice that the slopes
of the T-F relation and the $M_d vs. R_d $ relation almost exactly satisfy
equation (2), $ {3.5 \over 3.5-2} =2.33 $ which was derived merely on dimensional grounds, and
under the hypothesis of self similar halos having a constant baryon
fraction and constant $\lambda$. This result shows that as we are
assuming self similar halos, requiring that our final galaxies should
satisfy the two observational relations presented above will probably
yield galaxies with a constant $F$, and a constant $\lambda$, which is
what one might imagine would hold for a given class of galaxies.
 
At this point we have used all available observational constraints, and have to
introduce a fifth condition in order to fully fix our typical LSB
galaxy. The low surface brightness of these systems and the consequent
large scale radii imply a large angular momentum. Analytical (e.g. Peebles
1969, Catelan \& Theuns 1996) and numerical 
(e.g. Barnes \& Efstathiou 1987, Warren et al. 1992) studies of the
way tidal interactions between forming protogalactic fluctuations
transfer orbital angular momentum to rotational angular momentum in
these systems coincide in predicting values of
$\lambda$ for galaxies in the range 0.01 - 0.2, with a mean value
around 0.05. As discussed above, it is plausible to expect that LSB
galaxies are drawn from the high $\lambda$ region of the distribution,
so for our last restriction, we shall impose that this typical LSB
galaxy has a $\lambda=0.1$. In choosing a value for this parameter we 
are estimating the typical $\lambda$ of a high $\lambda$ system, there
can be little uncertainty in this number, perhaps of around a factor
of less than 1.5. This value will be our only free parameter, as
from this typical LSB we shall calibrate relations to use for other
galaxies.

Now that we have as many restrictions on our typical LSB
as initial conditions required for a model, we can
search the initial conditions space for the values of the five
parameters that result in the specified LSB. Given the shape of
the restrictions introduced, it is easy to see that this point will be
unique. The full rotation curve of this typical LSB is shown in Figure
(4) by the solid line. It can be seen that it resembles 
those of de Blok et al. (1996) rather well,
which is natural, as this is what was required of it. Figure (4) also
shows the final halo rotation curve (dotted line) which is seen to
form the dominant component in all but the extreme innermost
regions. The thin line in Figure (4) shows the rotation curve of the
initial King halo, which clearly did suffer some inward pull from the
formation of the disk. Figure (5) is analogous to Figure (4), but
shows the galaxy out to a much larger radial distance, where the drop
in the rotation curve is apparent, and an almost Keplerian region
appears outward of  30 kpc. It can also be seen that the adiabatic
response of the halo was confined to the inner regions.
Outside of $\approx 25 kpc$, only the subtraction of the baryon fraction from
the original halo is seen. 
We now look at the initial conditions which this model required.

\begin{figure}
\epsfig{file=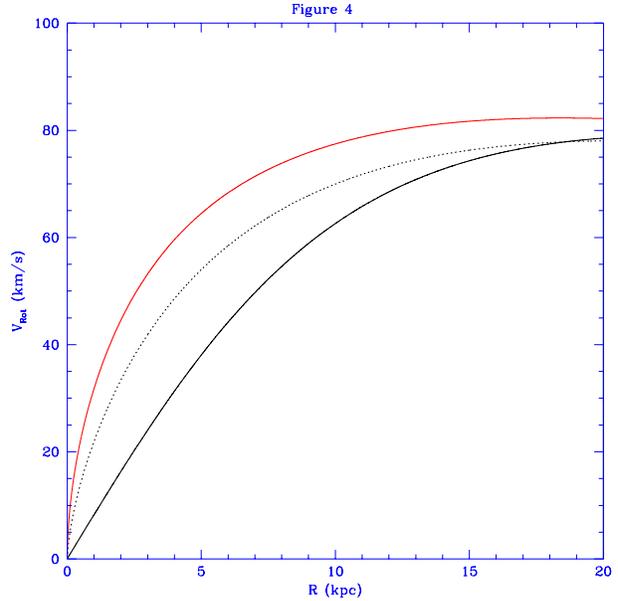,angle=0,width=8.5cm}
\caption{Final total rotation curve for the typical LSB modeled (solid line),
having $log(M)=10.72, log(M_d)=9.5$. Also
shown are the final and initial halo rotation curves, as dotted and thin lines, respectively.}
\label{Figure 4.}
\end{figure}

\begin{figure}
\epsfig{file=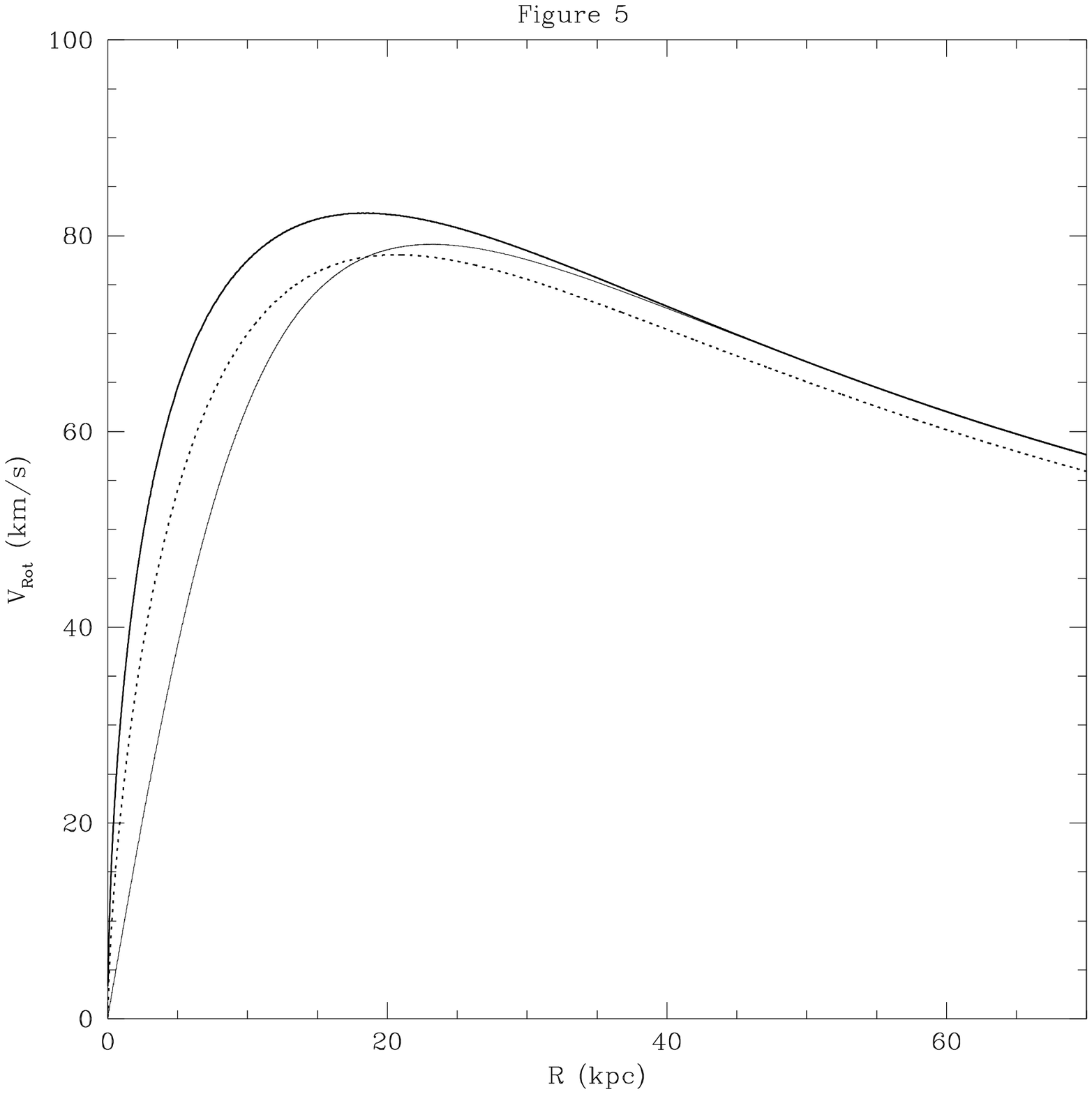,angle=0,width=8.5cm}
\caption{Same as Figure (4), but out to a larger radial distance.
Final total rotation curve for the typical LSB modeled (solid line),
having $log(M)=10.72, log(M_d)=9.5$. Also
shown are the final and initial halo rotation curves, as dotted and thin lines, respectively.}
\label{Figure 5.}
\end{figure}

\begin{figure}
\epsfig{file=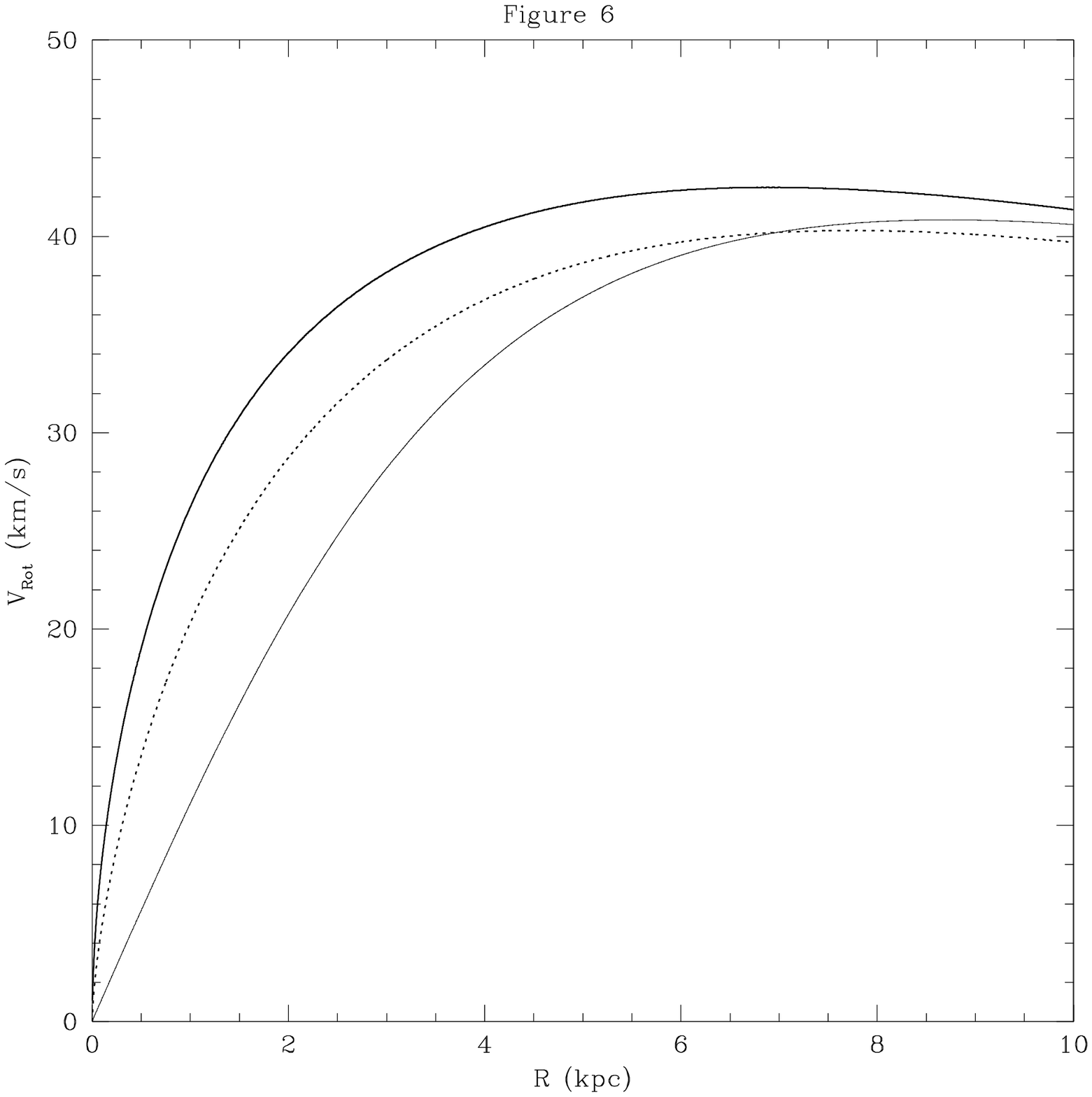,angle=0,width=8.5cm}
\caption{Final total rotation curve (solid line), and the final and initial
halo rotation curves as dotted and thin lines respectively, for a lower mass LSB galaxy,
having $log(M)=9.72$ and $log(M_d)=8.5$.}
\label{Figure 6.}
\end{figure}

\begin{figure}
\epsfig{file=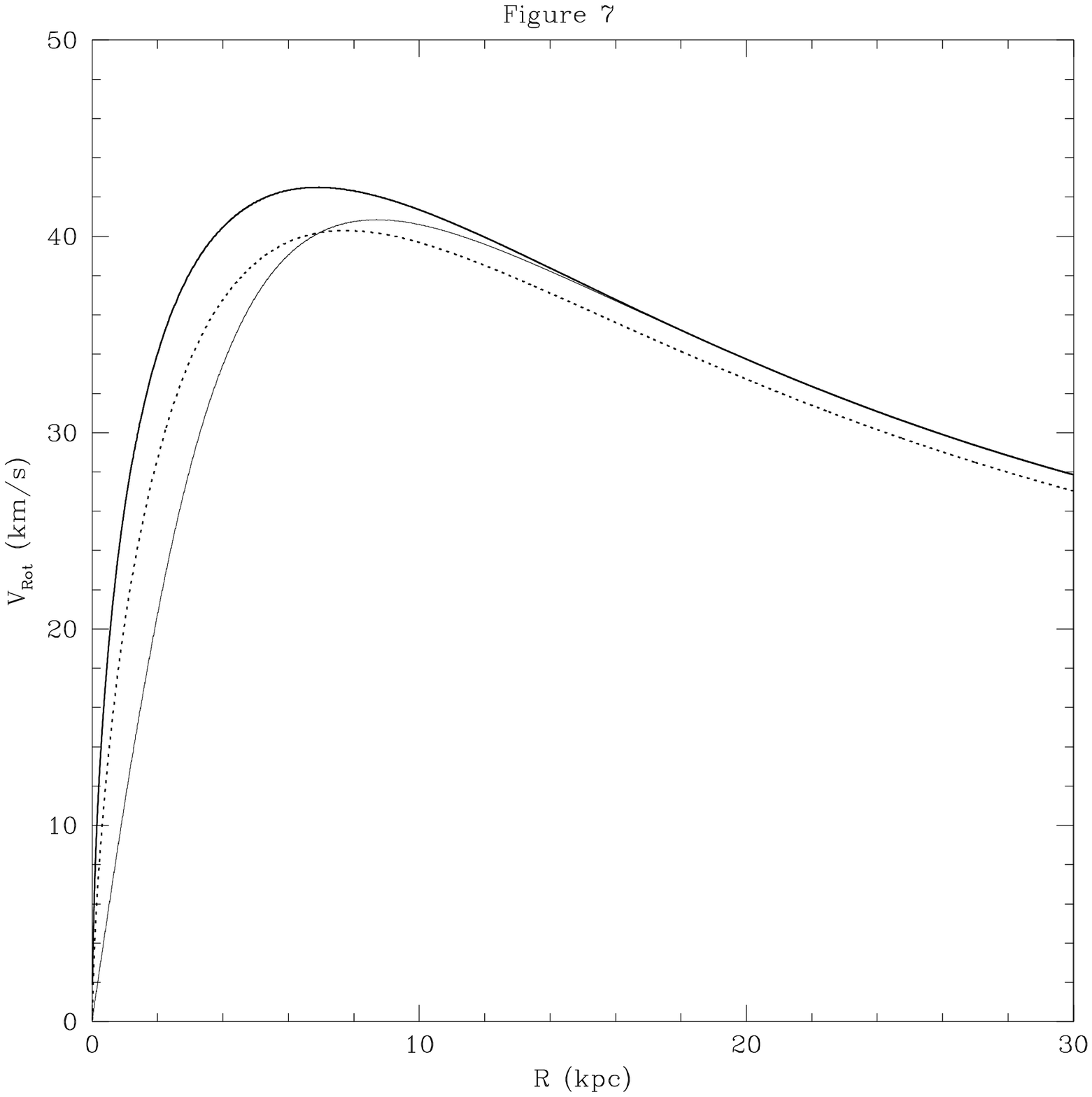,angle=0,width=8.5cm}
\caption{Same as Figure (6), but out to a larger radial distance.
Final total rotation curve (solid line), and the final and initial
halo rotation curves as dotted and thin lines respectively, for a lower mass LSB galaxy,
having $log(M)=9.72, log(M_d)=8.5$.}
\label{Figure 7.}
\end{figure}

\begin{figure}
\epsfig{file=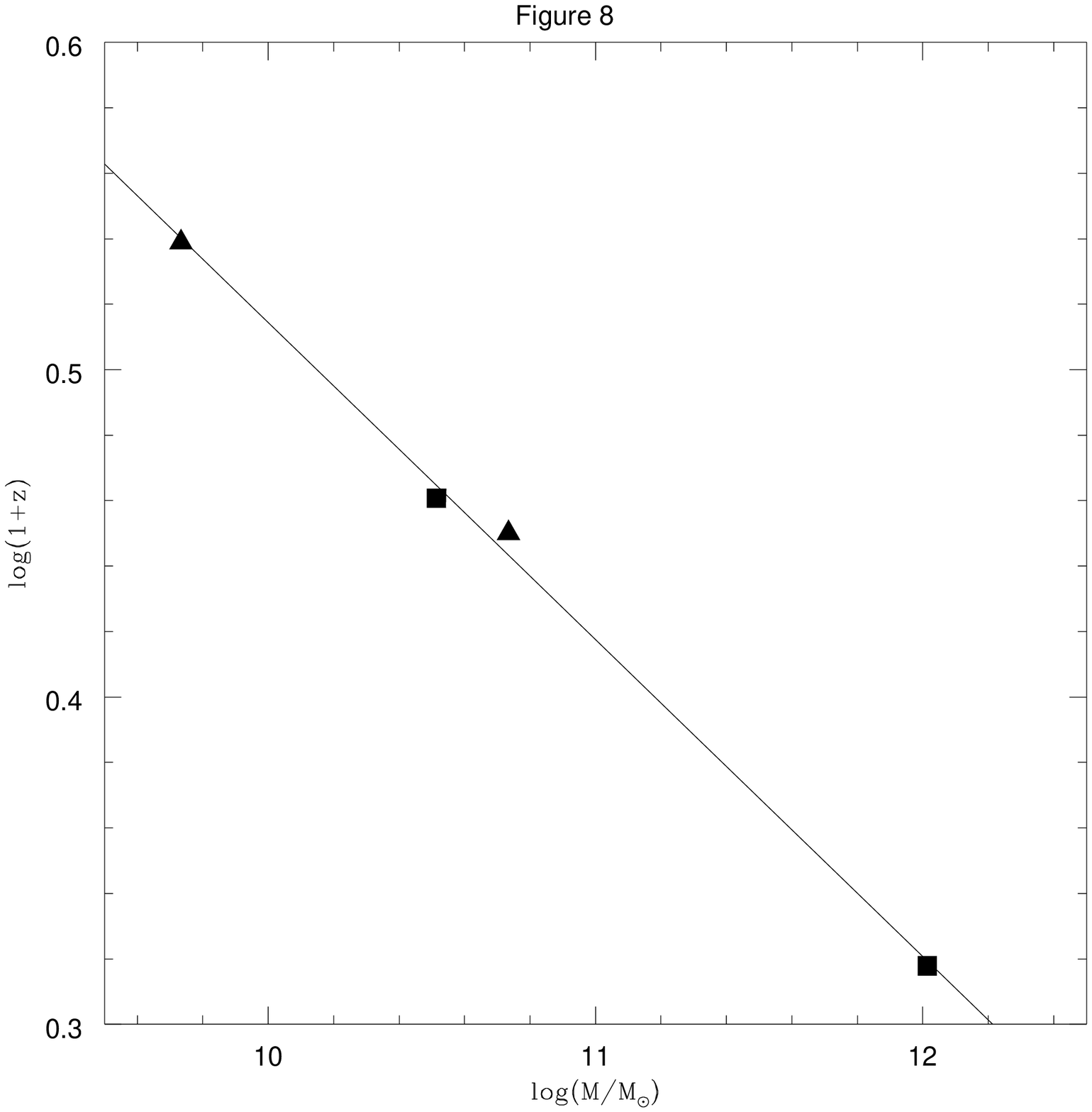,angle=0,width=8.5cm}
\caption{Plot of $log(1+z) vs. log(M)$ for the two LSB galaxies presented
and the two normal spirals (triangles and squares, respectively). The solid line is a linear
fit to a much larger sample of models of both classes.}
\label{Figure 8.}
\end{figure}

The shape parameter of the typical LSB is $P_0=4$, the baryon fraction is 
$F=0.06$, which gives $log(M)=10.72$ and is compatible with other estimates of
this number. The core radius of the initial King halo was 12.1 kpc, which
gives a total potential energy of $log(W)= -18.55 M_{\odot}(kpc/s)^2$ . In order to translate
this last number into something more recognizable, we shall use equation (6)
to associate a redshift with the formation of this system. This is done
only to have a more manageable quantity, and for comparisons between
models. Given that the formation history of the halos was not treated,
the values of redshift given are merely indicative. For
our typical LSB we obtain $z=1.8$. The value of $R_M$ was 44 kpc, not
in conflict with the rotation curves of these systems. Even though King
profiles have only a small region which approximates an isothermal
profile, after which an increasingly abrupt drop in the rotation curve 
follows, this region is of sufficient extent to reproduce the rotation
curve of these LSB galaxies. For normal spirals, where the flat
rotation curve region is much more extended, we shall find that the
greater relevance of the baryonic component in the inner regions
(which induces a much stronger contraction in the central regions of
the dark halo through the gravitational response associated with the
formation of the disk) is sufficient to give quite flat rotation
curves, which drop only beyond the observed regions. 

Taking $M$ as our independent parameter we now require 4 restrictions to
fully specify another galaxy. We shall use $P_0 =4$, $\lambda = 0.1$ and the
two observational constraint equations (10) and (11), to construct LSB
galaxies of different masses. Figures (6) and (7) are
analogous to Figures (4), (5), but for a galaxy total mass of $log(M)=9.72$. It
is interesting to see that for all these LSB galaxies of different
masses we obtain the same value for the baryon fraction of $F=0.06$,
which is certainly encouraging, as strong variations in this
parameter would be unexpected. The values of z obtained for this mass
sequence are shown in Figure (8), and are seen to lie along a
line with slope -0.097. This compares well with the prediction of eq(7) for
$\beta=2.33$ of -0.096. The agreement of these two independent approaches is encouraging. For
comparison, this yields $n=-2.4$, which agrees with independent
estimates of this number, despite the crudeness of the method used here.
Having taken self similar halos, and a
constant value of $\lambda$, all our LSB galaxies are self
similar in that they are scaled versions of each other. When compared on a linear scale, this last result
yields rotation curves which raise more gradually for larger 
galaxies, as seen in the sequence of Figures (4)-(7).

\subsection{The case of DDO 154} 
 
For the dwarf galaxy DDO 154 both the rotation curve and the light and
gas distributions have been accurately measured. This presents an ideal
opportunity to test the model, as the rotation velocity measurements
extend for over 20 disk scale radii, into the Keplerian region. This
is a clear case where the NFW profiles do not match the observation.
Burkert \& Silk (1997) show that if standard CDM (NFW) profiles are fitted to the
inner regions of this galaxy, they seriously over predict the outer
regions. The inner regions are similarly over predicted if a NFW
profile is fitted to the outer regions of this rotation
curve. Further, if the declining part of the rotation curve is fitted,
the inner regions are so strongly over predicted that it appears
unlikely that any secular processes in the minority baryonic component
could redistribute sufficiently large amounts of the dark halo profile
to match the observation. 
We fix $V_{M}=47.5 km/s$ from the observed rotation curve, $R_d$ 
from eq.(11) and $M_{d}=3.2 \times 10^{8} M_{\odot}$ from
the observed light and gas profiles, and use $P_0=4$ and $F=0.06$, as calibrated
from the LSB cases, to fully constrain this model.
The baryonic component in this galaxy is dominated by an extended gas
disk which is not exponential, but given this system is totally dark matter dominated, this does not 
influence the modeled rotation curve significantly, although it does make estimating the initial
value of $\lambda$ for this system rather model-dependent. The resulting final
rotation curve is shown in Figure (9), and is seen to reproduce
the data reasonably well at all radii, although the outer decreasing
region appears further out in the model. (Given that the observational relations used
to obtain $F$ and $P_0$ contain some spread, we could fine-tune the numbers to improve the
agreement. Further, there are observational uncertainties in the total gas and star content
and on the distance to this galaxy, all of which could be adjusted to present an optimized fit.
We do not do this as the spirit of our approach is its universality.) 
The modeled curve is thus not a fit, it is merely our standard LSB galaxy with the same
$V_M$ as DDO 154. We note that although only representative and central observational constraints
have been used in this analysis, the agreement with the whole rotation curve is quite good.
This, together with the morphological similarity of LSB and dwarf galaxies (de Blok et al 1996)
suggests that a halo profile of the form described here is universally applicable.

\begin{figure}
\epsfig{file=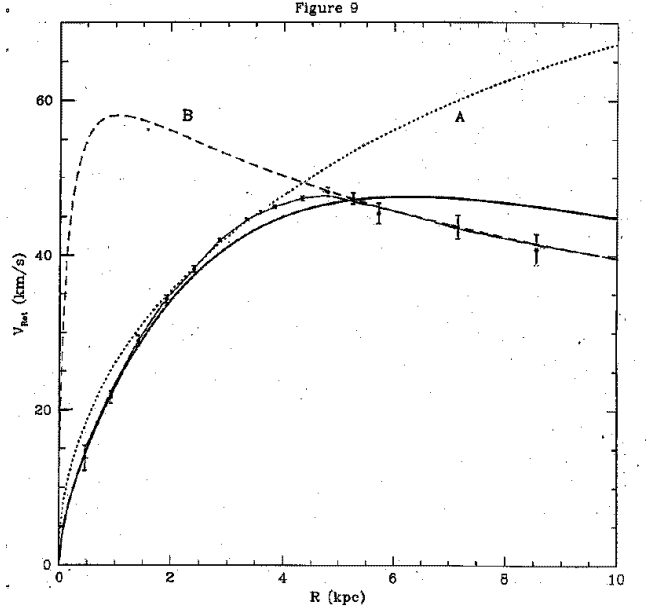,angle=0,width=8.5cm}
\caption{Rotation curve of DDO 154. The points with error bars
are the data from Carignan and Beaulieu (1997), reviewed in Burkert \& Silk (1997); 
the thin line is a fit to their data. Lines
A and B are NFW profiles fitted to the inner and outer regions, respectively, from
Burkert \& Silk (1997), and the thick line corresponds to our LSB galaxy of the required
$V_{M}$.}
\label{Figure 9.}
\end{figure}

\subsection{Direct profile comparisons}

We present a more explicit comparison between the King profile at $P_0$=4 with other
commonly used dark halo profiles in Figure (10). The thick solid curve in Figure (10)
is the final total rotation curve of our ``typical'' LSB galaxy, discussed in the previous section
and presented in Figures (4) and (5). This galaxy has an exponential disk of mass $log(M_d)=9.5$ with a
scale length as given by the data through eq.(11). The parameters of the initial King halo were
chosen to give the value of $V_M$ required by eq.(10), and a final central matter density consistent
with observations of LSB rotation curves, as explained previously. This procedure yielded a baryon 
fraction $F=0.06$. The dashed curve is the final total rotation curve calculated using the same
procedure as was used for the LSB galaxies, but starting off from a singular isothermal halo
profile. The final disk in this case is identical to that in the model corresponding to the thick solid curve,
and the isothermal halo was calibrated to give the same $V_M$. As can be seen, this halo is unable to 
reproduce the core region of the rotation curve seen in LSB galaxies, and produces a rotation curve
which remains essentially flat.
	The dotted curves represent galaxies starting off as Hernquist spheres (Hernquist 1990), 
$$
\rho (R_{H})=\left( {M \over {2\pi r_{H}^3}} \right) {1\over{R_{H}(R_{H}+1)^3}}
$$

where $R_H=r/r_H$, the radius in units of a scale radius for this profile and $M$ is the total
halo mass. The final disk was identical to that in the previous two cases. As this profile
has two independent parameters, the condition $V_M(M_d)$ of eq.(10) can be satisfied in any number
of ways, using Hernquist halos of different scale lengths and different total masses, but constant
maximum velocities (in combination with the fixed baryon disk). Taking $F=0.06$, as was found 
for the King spheres with $P_0=4$, we obtain the high central density Hernquist model (fastest rising
dotted curve). This model is again seen to fail in reproducing the low density core structure seen in the
rotation curves of LSB and dwarf galaxies. The central density of the Hernquist model can be reduced
taking a halo with a larger $r_H$, and hence a higher mass, as $V_M$ is kept fixed. Taking a halo
3 times as massive (keeping the final disk fixed) i.e. $F=0.02$ we obtain the low central density
Hernquist model, which still  has an excessively high central density, compared with the
King halo curve, which was modeled to reproduce the LSB curves. Additionally, such a low baryon
fraction seems unreasonable, and the high mass results in a rotation curve which remains flat 
out to much larger radial distances, which would be an additional problem in cases like DDO 154.

Finally, the thin continuous line represents a system having an initial profile given by

$$
\rho (R_{NFW})={\rho_{crit}\delta_{c} \over R(R+1)^2}
$$

where $R_{NFW}=r/r_{NFW}$, the radius in units of a scale radius, which is the fitting
formula which suitably approximates the dark halo profiles which arise in the self-consistent
cosmological simulations of Navarro et al. (1996b). $\rho_{crit}$ is $3H_{0}^2 / 8 \pi G$
and $\delta_c$ is a parameter given by the cosmology and $V_M$. In this case the two parameters of the
distribution are not independent, and once a cosmology is chosen (we took standard CDM and $h=0.75$:
standard CDM is the model which yields the lowest central density cores, Navarro et al. 1996b-their 
Figure 12), a relationship between them is fixed. Setting the total mass so that $V_M(M_d)$
agrees with eq.(10), we took the value of $\delta_c$ from their Figure 5 and from their Figure 6
the value of the constant used to calculate $r_{NFW}$. The final total rotation curve of this model is
essentially equivalent to the one of the high central density Hernquist model, and similarly
inconsistent with the core region inferred from the rotation curves discussed above, as pointed
out in the previous subsection.

These comparisons have been made including the final baryon disk and the adiabatic response of the
initial halo to the disk formation not only to establish a common standard for the comparison
(a fixed final disk) but also to illustrate to what extent the observed rotation curves of
LSB and dwarf galaxies serve to discriminate between different halo profiles. As we took the highly
extended disks of LSB galaxies, the distortion to the original halo is minimized, and the systems remain
dark matter dominated. It can be seen that the low density core region seen in the observed rotation 
curves can not be reproduced using a halo profile which is centrally divergent, as with the 3 
we have examined here.

\section{Normal late type galaxies and comparisons}

In this section we shall use the method described above to estimate the
initial formation conditions of a mass sequence of typical late type
galaxies. To do this we use the halo shape $P_0=4$ which results from
fitting the LSB galaxies. As a second restriction, we use
$F=0.06$, also obtained from the LSB galaxies, as we speculate that the
main difference between LSB and other late type galaxies is the value
of $\lambda$. No variations in $F$ with mass were found for the LSB
sequence, so we expect any mass loss in these systems has been marginal,
consistent with their large potential depths. Equation (10) is used for the T-F
relation, assumed to hold for normal late type spirals. As a
final restriction, we take the list of total luminosities and disk
scale radii given at the end of Persic et al. (1996) (taking only galaxies with T$>4$) to produce
Figure (11). A linear regression to these data gives:

\begin{equation}
log(M_d) = 9.33 + 2.33log(R_d).   
\end{equation}

\begin{figure}
\epsfig{file=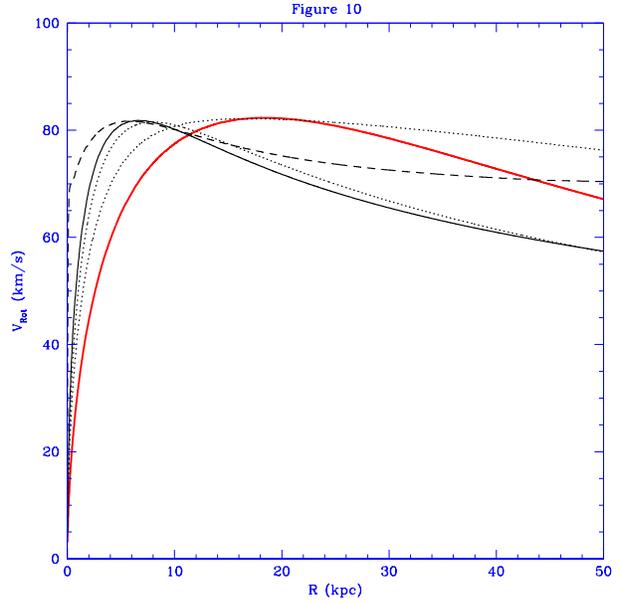,angle=0,width=8.5cm}
\caption{Final total rotation curve for the typical LSB modeled 
having $log(M)=10.72, log(M_d)=9.5$, for different choices of the initial halo profile.
The thick solid line corresponds to the King profile of Figures (4), (5),
the dashed line assumes a singular isothermal halo. The dotted lines are two
different Hernquist models, having baryon fractions of 0.06 (higher central density
) and 0.02 (lower central density). The thin solid line was calculated using the
only NFW profile having the required $V_M$ for the cosmology which gives the
lowest central density NFW profiles.}
\label{Figure 10.}
\end{figure}

\begin{figure}
\epsfig{file=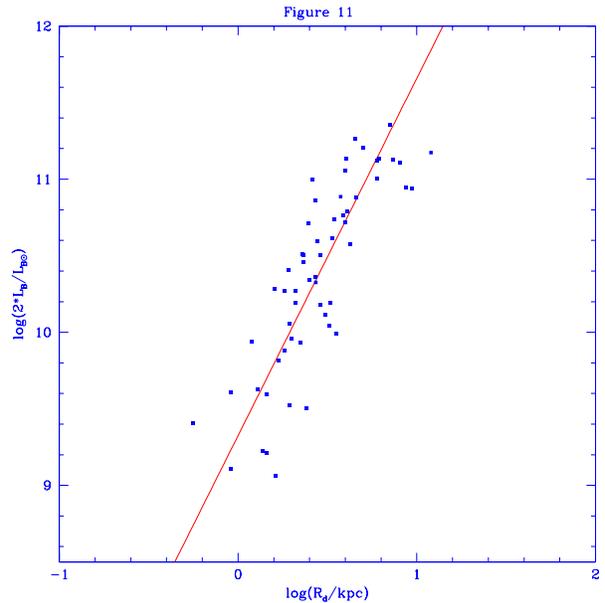,angle=0,width=8.5cm}
\caption{Log $(M_d)$ vs. log$(R_d)$ for the T$>4$ normal spirals listed in
Persic et al. (1996), assuming a $M_{\odot}/L_{B\odot}$ ratio of 2. Also shown is a linear
fit to the data. The slope of this line
is the same as that in Figure(3), which contains the analogous information for the LSB galaxies.}
\label{Figure 11.}
\end{figure}

Equation (12) has essentially the same slope as the analogous relation for the LSB
galaxies, but for a given $M_d$, predicts a scale radius 2.7 times
smaller. Thus, this observed scaling relation is also consistent with the
observed T-F in the way the dimensional analysis would imply.
At this point we can generate a mass sequence of late type spirals,
which satisfy the 4 conditions above at every mass, obtain the full
rotation curves and solve for $\lambda$, $z$ and $R_M$, which can then be
compared to the values obtained for the LSB galaxies. 
Note however that these galaxies are very baryon-dominated in their inner regions,
where our model is unlikely to work well. Our aim here is to show that our halo
model is not inconsistent with observations in these cases, not to obtain
exact agreement with data.

\begin{figure}
\epsfig{file=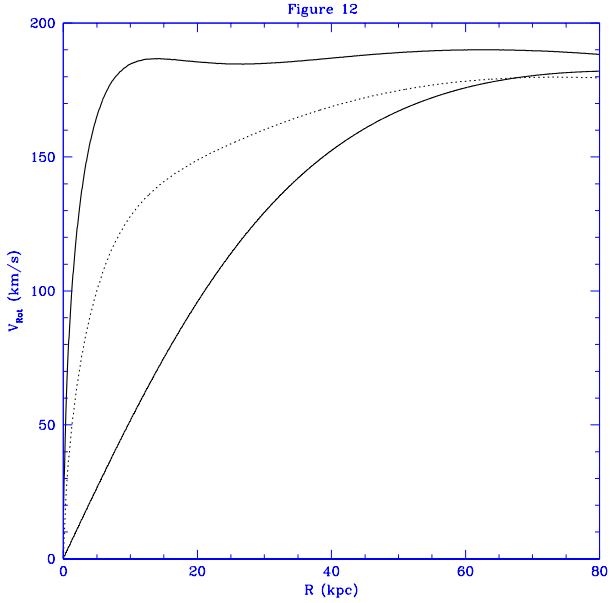,angle=0,width=8.5cm}
\caption{Final total rotation curve (solid line). Also
shown are the final and initial halo rotation curves, as dotted and thin lines respectively,
for a log($M$)=12 normal spiral. This can be compared with Figure (4) for LSB galaxies.}
\label{Figure 12.}
\end{figure}

\begin{figure}
\epsfig{file=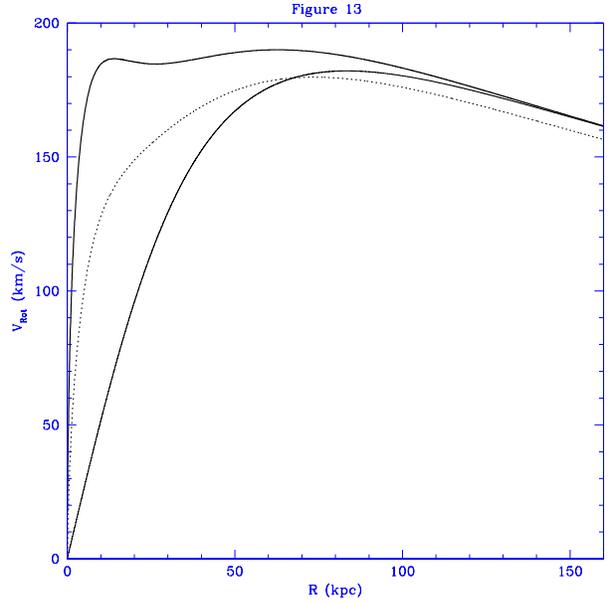,angle=0,width=8.5cm}
\caption{Same as Figure (12), but out to a larger radial distance.
Final total rotation curve (solid line). Also
shown are the final and initial halo rotation curves, as dotted and thin lines respectively,
for a log($M$)=12 normal spiral.}
\label{Figure 13.}
\end{figure}

Figure (12) shows the total final rotation curve, solid line, final
halo rotation curve, dotted line, and initial halo rotation curve for
a $log(M)=12$ normal late type spiral, out to 80 kpc, Figure (13) is
analogous, but out to 160 kpc. It can be seen that this rotation
curves closely match observed ones in general shape. The decreasing
region is not apparent until after 100 kpc. Considering an extra bulge component
would improve the agreement with observations, as a central overshooting would be produced,
but this has not been done to avoid new parameters. Given the value of $F$
taken, this model should approximate our galaxy and has a $V_M=190
kms^{-1}$, rather than 220 as the observed one. This may be a result of 
our galaxy lying slightly above of the mean T-F relation. A slight 
change in the $M/L$ for our galaxy would also furnish agreement.
Either way this discrepancy is not surprising given the
approximations of the method, which is being applied in a case where the
assumptions become less plausible. Notice that in this case the flat region
of the rotation curve is much more extended than in the LSB cases, and that the halo has been
distorted to a larger degree, the baryons having contracted much
more. For this model we get a value of $\lambda=0.04$, which is quite
close to the mean in the predicted $\lambda$ distributions, and gives
some confidence on the value of 0.1 used for the LSB's. As was
expected, we obtain smaller values of $\lambda$ than for the
LSB's. 

\begin{figure}
\epsfig{file=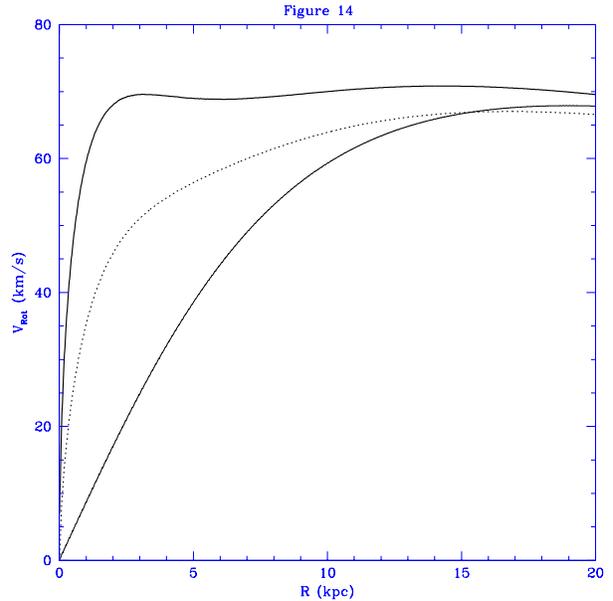,angle=0,width=8.5cm}
\caption{Final total rotation curve (solid line). Also
shown are the final and initial halo rotation curves, as dotted and thin lines respectively,
for a log($M$)=10.5 normal spiral.}
\label{Figure 14.}
\end{figure}

\begin{figure}
\epsfig{file=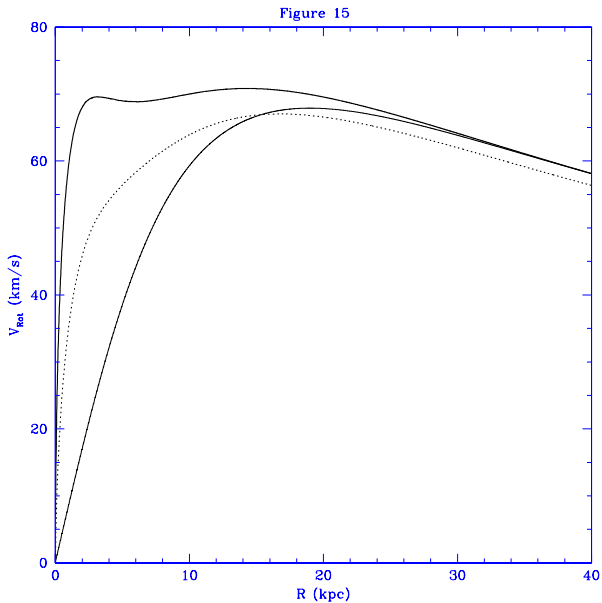,angle=0,width=8.5cm}
\caption{Same as Figure (14), but out to a larger radial distance.
Final total rotation curve (solid line). Also
shown are the final and initial halo rotation curves, as dotted and thin lines respectively,
for a log($M$)=10.5 normal spiral.}
\label{Figure 15.}
\end{figure}

For the rest of the mass sequence we obtain identical values of
$\lambda$, which is what the analytic results of section (2) led us to
expect. Here again having taken self similar halos, and obtaining
constant $\lambda$ for the normal late type sequence, we obtain self
similar final rotation curves, which are all scaled versions of each
other. Figures (14) and (15) are analogous to (12) and (13), but for a
total mass of $log(M)=10.5$. The description of the previous
model applies here. 

The values of $z$ for this mass sequence are the same as were found for the
LSB galaxies and are also shown in Figure (8), i.e. we do not find a significant
differences in the formation epochs of these two galaxy types. The blue
colors and large gas fractions of LSB galaxies are better explained in
terms of slow evolution timescales, in this case naturally arising from the low
surface density (Firmani \& Tutukov 1994, Firmani et al 1996), and not as reflecting recent
formation epochs.

Figure (16) shows $R_M vs. M$ for the normal spirals and
for the LSB sequence. The modeled galaxies lie on a line of slope 0.46. The arguments of
section (2) would suggest $R_{M}=R_{d}/\lambda$ $\Rightarrow R_{M} \propto M^{1/\beta}$
i.e., a slope of 0.43. Again, the agreement of the two independent approaches is encouraging.
It can be seen that the decrease in
the rotation curve appears at the same relative radial distances for
both types of galaxies. Thus a relation of this nature is a robust
prediction of this approach, awaiting more cases like DDO 154, which
lies at the cross in Figure (16) (an error in the adopted distance of a factor of 1.4 would make it coincide
with the models).

Figure (17) is analogous to Figure (12), but shows the rotation curves
in a log-log plot, where the characteristic shapes can be
better appreciated. It is clear that the baryonic component dominates
interior to around 10kpc, and the dark component outwards of this.
It can be seen that both the baryonic component
and the dark halo reaction to the disk formation produce an
enhancement in the total rotation curve slightly interior to the
original maximum, and of similar magnitude. This produces a large
``flat'' region, which has been identified as the signature of an
isothermal halo structure. In our model, 
this coincidence is tied to $F \approx \lambda$,
as the baryonic fraction $F$ contracts by a factor of $\approx 1/\lambda$,
hence matching the DM contribution in the inner regions. In this scaling, all other late type spiral
rotation curves we obtain appear as translations of the one shown.

Figure (18) is equivalent to (17), but shows the results for an LSB
galaxy, again, all other LSB galaxies would appear as
translations. Here the dominance of the baryonic component starts much
further in than in the previous case, and never reaches the same
extent. In this second case, it can be seen that the much smaller contraction of
the baryons induces a correspondingly milder reaction in the dark
halo. These last two effects are not sufficient to produce the large
flat region seen in the normal spirals. Formation of the disk does
however modify the inner regions of the rotation curve to some extent,
as can be seen in the differences between the 3 curves,
particularly in the inner 3 kpc. 

In both the normal spirals and LSB's
the halo reaction is confined to the interior 12 kpc, where the
concentration of the small fraction of baryons is sufficient to become
important. Exterior to this region in both cases, the rotation curves
are totally dark matter dominated, and not significantly affected by the disk
formation. In particular the total extent of the halo is not affected by disk formation.
This last result indicates that measuring the outer,
decaying portion of galactic rotation curves will furnish
much more strict restrictions on halo shapes and galactic formation
theories than observations of inner regions.

From the comparison of the log-log plots of Figures (17) and (18) it can be seen that  
disk formation and acompaning halo reaction produce an enhancement in the rotation 
curve interior to the original maximum, and only up to the same amplitude.
The ``asymptotic'' value is not affected. This last point is very
important as it shows that at a fixed point in the T-F diagram, taking a galaxy with a
different disk scale length alters only the model $\lambda$, as we saw that both galaxy types lie 
on the same $z(M)$ relation. 

In terms of the initial conditions, $\lambda$ controls $R_d$ but does not affect
$V_M$, while $z(M)$ determines $V_M$ and also has an effect on $R_d$, as can be seen from the dimensional analysis
in section 2. The dispersion in the T-F used here is 0.2 in $log(M)$ at fixed $V_M$. This introduces a
dispersion of 0.1 in $log(z)$ and of 0.2 in $log(R_d)$, which is sufficient to explain the 
dispersion in $R_d$ seen in the observed relations for both LSB and normal late type galaxies. 
However, it is rather contrived to think of a bimodal $P(\lambda)$. Rather it would appear
that selecting only a fixed galaxy class yields results compatible with a constant
$\lambda$. If we take the dispersion in $R_d$ at fixed $M$ as going from one limit of the
normal late type galaxies to the other limit of LSB galaxies, a total dispersion of almost
one order of magnitude appears. 

As a dispersion of only 0.2 in $log(R_d)$ is compatible with
$\lambda=constant$ , the dispersions in the observed galactic parameters imply we are seeing two
independent distributions, $P(z;M)$ and $P(\lambda)$, both of which have an intrinsic
dispersion, about 0.1 in $log(z)$ and at least 0.7 in $log(\lambda)$. The dispersion in $\lambda$ 
could be larger, as earlier type galaxies perhaps have even lower values, while more extended
LSB galaxies of lower surface brightness (even ``failed galaxies'') could still linger
undetected. Systematic variations in $F$ could exist, and complicate the above picture, but without any
compelling observational or theoretical reason to include them, this shall not be discussed here. 
Any dispersions in $F$, $M/L$ ratios or the introduction of efficiency factors would
reduce the scatter in $z(M)$ and $\lambda$ we deduce above as resulting from the observed dispersion
in $R_d$ and $V_M$, so that the values given are probably upper limits in this sense.

Although the choice of $\lambda_{LSB} =0.1$ (the value of $\lambda$ for LSB galaxies) yielded a 
calibration of the model which proved quite successful in giving a reasonable $F=0.06$, $\lambda_{NS}=0.04$,  
(the value of $\lambda$ for normal late type spirals), general shape of the rotation curves
of LSB and late type galaxies, and fair agreement with the rotation curve of DDO 154, we can not be 
certain of this value. Taking different values for our parameter within the range $0.07<\lambda<0.15$
leaves all our qualitative results unchanged, we have still found initial conditions such that the final
results match the observations adequately. Within this range $F$ and $\lambda$ scale linearly with
$\lambda_{LSB}$, i.e. $\lambda_{LSB}=0.15 \Longrightarrow \lambda_{NS}=0.06, F=0.09$, as could
be expected from the previous discussion. Accordingly, $z(M)$ and $R_{M}$ are not affected, neither is $P_{0}$.
The plot of $R_M$ is thus a robust prediction of this approach, albeit one which will be difficult to 
verify.

\begin{figure}
\epsfig{file=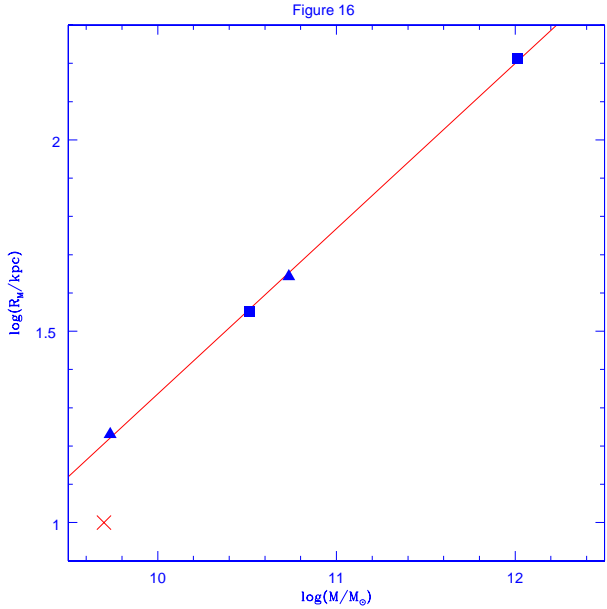,angle=0,width=8.5cm}
\caption{Plot of $R_{M} vs. M$ for the two LSB galaxies presented
and the two normal spirals (triangles and squares respectively). The solid line is a linear
fit to a much larger sample of models of both classes. The cross marks the position of DDO 154
(see text).}
\label{Figure 16.}
\end{figure}

\begin{figure}
\epsfig{file=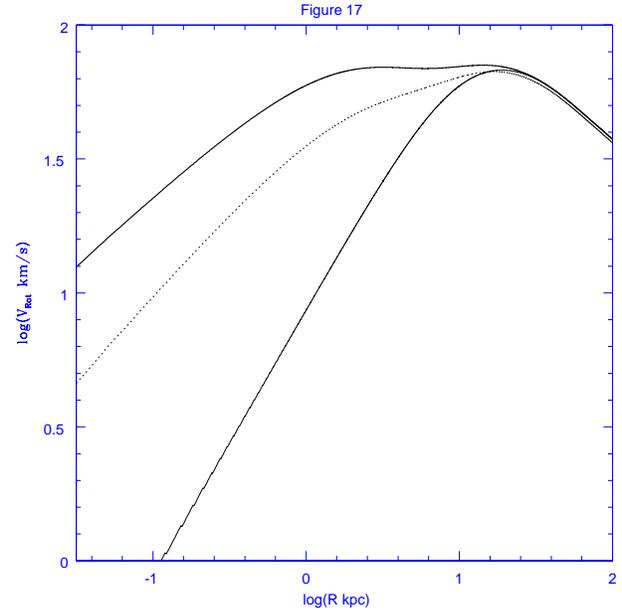,angle=0,width=8.5cm}
\caption{This figure shows the final total rotation curve 
(thick line) and the final and initial halo rotation curves (dotted and thin
lines) on a log-log scale,
for the normal spirals. All of our normal spirals appear as translations of 
each other in this plot, which is intended to highlight the intrinsic shape
of this curves.}
\label{Figure 17.}
\end{figure}

\begin{figure}
\epsfig{file=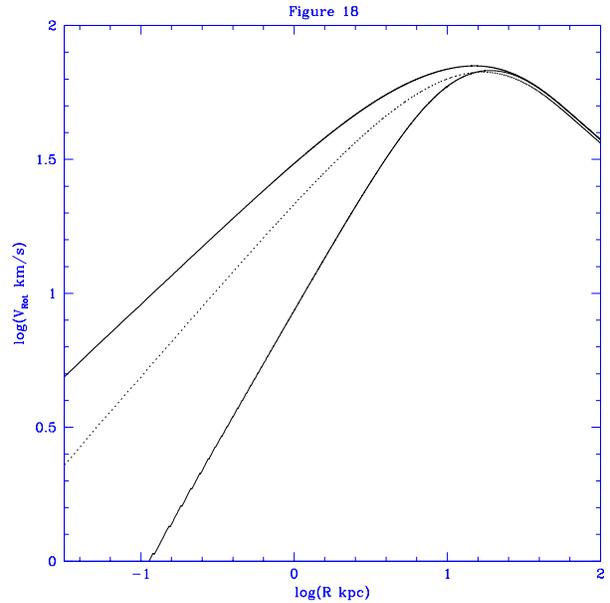,angle=0,width=8.5cm}
\caption{This figure shows the final total rotation curve 
(thick line) and the final and initial halo rotation curves (dotted and thin
lines) on a log-log scale, for our modeled LSB galaxies. Notice the milder
halo reaction to the disk formation, and the narrower 'flat' region.}
\label{Figure 18.}
\end{figure}

\section{Conclusions}

From the observational data shown, the theoretical approach presented and
the results obtained, we conclude the following:

1) The self similar initial halos of late type galaxies can be accurately
modeled as King halos with a shape parameter $\Phi_{0}/\sigma^2 =4$, the 
gravitational potential at the centre in units of the halo velocity
dispersion.

2) The baryon content of late type galaxies is of $0.065 \pm 0.025$.

3) The value of $\lambda$ (the characteristic angular momentum)
for normal type spirals is $0.045 \pm 0.015$.
A scaling between the galactic disc scale radius, $R_d$, and the
disk mass, $M_d$, of the form
$R_{d} \propto M_{d}^{2.33}$ represents curves of $\lambda =cte.$

4) The Tully-Fisher law reflects $z(M)$, the relation between the total halo mass, $M$,
and the halo potential energy, expressed through a nominal formation redshift, $z$.
LSB and normal late type spirals have the same $z(M)$, there is no
indication of a different formation epoch.
The total extent of late type galaxies halos scales as $R_{M} \propto M^{0.43}$,
where $R_M$ is a measure of the total extent of the halo, here the radius at which
the rotation curve has fallen to 85

5) $P(z;M)$ and $P(\lambda)$ are two independent relations, having dispersions of
around 0.1 and 0.7 in the log, respectively.

\section*{Acknowledgments}

The work of X. Hernandez was partly supported by a DGAPA-UNAM grant.

\end{document}